\documentclass[aps,prl,reprint,twocolumn,superscriptaddress,floatfix,nofootinbib,longbibliography]{revtex4-1}
\usepackage{graphicx,amsmath,amsfonts,amssymb,amsthm,xr}
\usepackage{epsfig,amsmath,amssymb,color,dsfont,upgreek,physics}
\usepackage{mathrsfs}
\usepackage{widetable}
\usepackage{mathtools}
\usepackage{bbold}
\usepackage{comment}
\usepackage{float}

\usepackage[bookmarks=true,colorlinks,linkcolor=OrangeRed,urlcolor=NavyBlue,citecolor=RoyalBlue]{hyperref}
\usepackage[dvipsnames]{xcolor}
\usepackage{orcidlink}

\definecolor{mygold}{rgb}{0.93,0.69,0.13}

\definecolor{mypurple}{rgb}{0.49,0.18,0.56}

\definecolor{mygreen}{rgb}{0,0.5,0}

\definecolor{mygreen}{rgb}{0,0.5,0}
\definecolor{myred}{rgb}{0.7,0,0}
\definecolor{myblue}{rgb}{0,0,1}

\begin{document}
\title{Observation of microscopic confinement dynamics by a tunable topological $\theta$-angle}

\author{Wei-Yong Zhang}
\thanks{W.-Y.Z.~, Y.L.~and Y.C~contributed equally to this work.}
\affiliation{Hefei National Research Center for Physical Sciences at the Microscale and School of Physics, University of Science and Technology of China, Hefei 230026, China}

\author{Ying Liu}
\thanks{W.-Y.Z.~, Y.L.~and Y.C~contributed equally to this work.}
\affiliation{Hefei National Research Center for Physical Sciences at the Microscale and School of Physics, University of Science and Technology of China, Hefei 230026, China}

\author{Yanting Cheng}
\thanks{W.-Y.Z.~, Y.L.~and Y.C~contributed equally to this work.}
\affiliation{Institute of Theoretical Physics and Department of Physics, University of Science and Technology Beijing, Beijing 100083, China}

\author{Ming-Gen He}
\affiliation{Hefei National Research Center for Physical Sciences at the Microscale and School of Physics, University of Science and Technology of China, Hefei 230026, China}

\author{Han-Yi Wang}
\affiliation{Hefei National Research Center for Physical Sciences at the Microscale and School of Physics, University of Science and Technology of China, Hefei 230026, China}

\author{Tian-Yi Wang}
\affiliation{Hefei National Research Center for Physical Sciences at the Microscale and School of Physics, University of Science and Technology of China, Hefei 230026, China}

\author{Zi-Hang Zhu}
\affiliation{Hefei National Research Center for Physical Sciences at the Microscale and School of Physics, University of Science and Technology of China, Hefei 230026, China}

\author{Guo-Xian Su}
\affiliation{Hefei National Research Center for Physical Sciences at the Microscale and School of Physics, University of Science and Technology of China, Hefei 230026, China}

\author{Zhao-Yu Zhou}
\affiliation{Hefei National Research Center for Physical Sciences at the Microscale and School of Physics, University of Science and Technology of China, Hefei 230026, China}

\author{Yong-Guang Zheng}
\affiliation{Hefei National Research Center for Physical Sciences at the Microscale and School of Physics, University of Science and Technology of China, Hefei 230026, China}

\author{Hui Sun}
\affiliation{Hefei National Research Center for Physical Sciences at the Microscale and School of Physics, University of Science and Technology of China, Hefei 230026, China}

\author{Bing Yang}
\affiliation{Department of Physics, Southern University of Science and Technology, Shenzhen 518055, China}

\author{Philipp Hauke}
\affiliation{INO-CNR BEC Center and Department of Physics, University of Trento, Via Sommarive 14, I-38123 Trento, Italy}
\affiliation{INFN-TIFPA, Trento Institute for Fundamental Physics and Applications,Via Sommarive 14, Trento, I-38123, Italy}

\author{Wei Zheng}
\affiliation{Hefei National Research Center for Physical Sciences at the Microscale and School of Physics, University of Science and Technology of China, Hefei 230026, China}
\affiliation{CAS Center for Excellence in Quantum Information and Quantum Physics, University of Science and Technology of China, Hefei 230026, China}
\affiliation{Hefei National Laboratory, University of Science and Technology of China, Hefei 230088, China}

\author{Jad C. Halimeh}
\affiliation{Department of Physics and Arnold Sommerfeld Center for Theoretical Physics (ASC), Ludwig-Maximilians-Universit\"at M\"unchen, Theresienstraße 37, D-80333 M\"unchen, Germany}
\affiliation{Munich Center for Quantum Science and Technology (MCQST), Schellingstraße 4, D-80799 M\"unchen, Germany}

\author{Zhen-Sheng Yuan}
\affiliation{Hefei National Research Center for Physical Sciences at the Microscale and School of Physics, University of Science and Technology of China, Hefei 230026, China}
\affiliation{CAS Center for Excellence in Quantum Information and Quantum Physics, University of Science and Technology of China, Hefei 230026, China}
\affiliation{Hefei National Laboratory, University of Science and Technology of China, Hefei 230088, China}

\author{Jian-Wei Pan}
\affiliation{Hefei National Research Center for Physical Sciences at the Microscale and School of Physics, University of Science and Technology of China, Hefei 230026, China}
\affiliation{CAS Center for Excellence in Quantum Information and Quantum Physics, University of Science and Technology of China, Hefei 230026, China}
\affiliation{Hefei National Laboratory, University of Science and Technology of China, Hefei 230088, China}

\begin{abstract}
    The topological $\theta$-angle is central to the understanding of a plethora of phenomena in condensed matter and high-energy physics such as the strong CP problem \cite{Mannel2007}, dynamical quantum topological phase transitions \cite{Zache2019dynamical,Huang2019dynamical}, and the confinement--deconfinement transition \cite{Gross1996,Buyens2016confinement,Surace2020lattice}. Difficulties arise when probing the effects of the topological $\theta$-angle using classical methods, in particular through the appearance of a sign problem in numerical simulations \cite{unsal2012theta,banuls2020review,Berges2021qcd}. Quantum simulators offer a powerful alternate venue for realizing the $\theta$-angle, which has hitherto remained an outstanding challenge due to the difficulty of introducing a dynamical electric field in the experiment \cite{Schweizer2019,Mil2020,Yang2020observation}. Here, we report on the experimental realization of a tunable topological $\theta$-angle in a Bose--Hubbard gauge-theory quantum simulator, implemented through a tilted superlattice potential that induces an effective background electric field. We demonstrate the rich physics due to this angle by the direct observation of the confinement--deconfinement transition of $(1+1)$-dimensional quantum electrodynamics. Using an atomic-precision quantum gas microscope, we distinguish between the confined and deconfined phases by monitoring the real-time evolution of particle--antiparticle pairs, which exhibit constrained (ballistic) propagation for a finite (vanishing) deviation of the $\theta$-angle from $\pi$. Our work provides a major step forward in the realization of topological terms on modern quantum simulators, and the exploration of rich physics they have been theorized to entail.
\end{abstract}

\date{\today}
\maketitle

\textbf{\textit{Introduction.---}}Gauge theories are a fundamental framework in modern physics that encodes the laws of nature through a local symmetry. This gauge symmetry induces an extensive number of local constraints between matter and gauge fields, a prime example being Gauss's law in quantum electrodynamics (QED) that couples electrons and positrons with the electromagnetic field at each point in space \cite{Weinberg_book}. A prominent feature that naturally occurs in certain gauge theories due to the topological nature of the vacuum is the topological $\theta$-term \cite{Jackiw1976,Callan1976,tHooft1976}. This term is responsible for the strong CP problem in $3+1$D quantum chromodynamics (QCD), a gauge theory describing the strong force \cite{Mannel2007}. In $1+1$D QED, the topological $\theta$-term can induce a transition between confined and deconfined phases \cite{Buyens2016confinement,Surace2020lattice}. This topological term can also emerge in low-energy effective theories of condensed matter, such as axion electrodynamics in topological insulators and fractional excitations in spin chains \cite{li2010dynamical,haldane1983continuum,haldane1983nonlinear}.

Recently there has been a surge of interest in the quantum simulation of gauge theories \cite{Dalmonte2016lattice,Zohar2016quantum,Banuls2020simulating,Aidelsburger2022cold,Zohar2022quantum}, providing a means of probing high-energy questions that is complementary to dedicated setups such as particle colliders and classical high-performance computing.
The key to quantum simulating a large-scale gauge theory is the ability to stabilize the gauge symmetry reliably \cite{Halimeh2020reliability}. 
In the last years, there have been remarkable breakthroughs to overcome this formidable challenge 
\cite{Martinez2016,Schweizer2019,Goerg2019,Mil2020,Yang2020observation,Zhou2022thermalization,Nguyen2021,Wang2021,Mildenberger2022}. Nevertheless, despite impressive recent progress, the realization of the topological $\theta$-term has remained elusive due to the difficulty of realizing a dynamical electric field in the experiment \cite{Yang2020observation}.

\begin{figure*}[htb]
    \centering
    \includegraphics[width=0.75\textwidth]{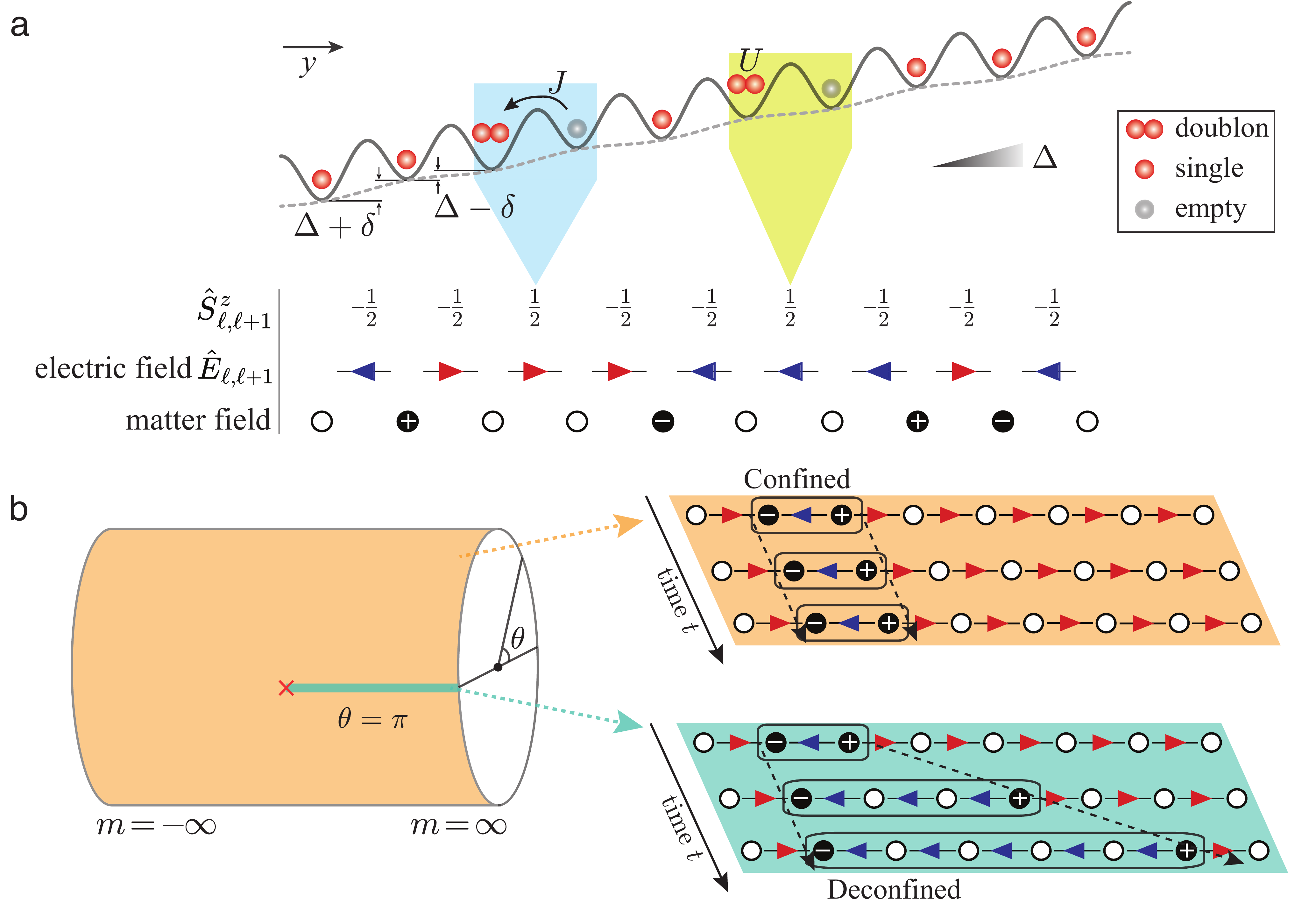}
    \caption{\textbf{Experimental implementation of the $\mathrm{U}(1)$ QLM and sketch of its phase diagram.}
    (a) The physical model for bosons in a one-dimensional optical superlattice in the presence of a linearly tilted potential (see the mapping in Table \ref{tab:mapping}). The local electric field is given by $\hat{E}_{\ell,\ell+1}^z=(-1)^{\ell+1}\hat{S}_{\ell,\ell+1}^z$ (indicated by arrows), from which the charge configuration can be read off thanks to the Gauss's law. The symbols {\small $\bigcirc$}, \mbox{\raisebox{-0.7ex}{\huge $\bullet$}{\hspace{-0.95em}\textcolor{white}{$+$}}}, and \mbox{\raisebox{-0.7ex}{\huge $\bullet$}{\hspace{-0.95em}\textcolor{white}{$-$}}} denote electric charge zero, $+1$, and $-1$ at the matter sites of the QLM.
    (b) Phase diagram of the $\mathrm{U}(1)$ QLM and the schematic illustrations of microscopic dynamics for the (de)confined phases. 
    \textbf{Left}: For $\theta=\pi$ ($\delta=0$), the system hosts a global $\mathbb{Z}_2$ translation symmetry and there is a confinement--deconfinement transition at $m = m^{\star}$ (the critical point, red cross marker). For $\theta\neq\pi$ ($\delta\neq0$), the $\mathbb{Z}_2$ translation symmetry is broken, and the system is in the confined phase.
    \textbf{Bottom right}: In the deconfined phase, the members of the particle--antiparticle pair propagate ballistically away from each other. The distance between them increases linearly with the evolution time.
    \textbf{Top right}: In the confined phase, the distance between the particle--antiparticle pair is energetically costly, and the pair is bound.
    }
    \label{figure1:models}
\end{figure*}

\begin{table*}[ht!]
    \caption{Mapping between the QLM and the BHM and implementing the topological $\theta$-angle.}
    \label{tab:mapping}    
    \centering
    \begin{widetabular}{\textwidth}{c|c|c}
    \hline\hline
    QLM & Mapping parameters & BHM  \\ 
    \hline 
    \parbox[h]{0.32\textwidth}{\vspace{-5pt} \begin{align}\nonumber 
    \hat{H}_{\mathrm{QLM}} = & -\kappa \sum_{\ell=1}^{L-1} \left( \hat{\psi}_\ell \hat{S}_{\ell,\ell+1}^{+} \hat{\psi}_{\ell+1} + \text{H.c.}\right) \\ \nonumber
    & + m \sum_{\ell=1}^L \hat{\psi}_\ell^{\dagger} \hat{\psi}_\ell \\ \nonumber
    & + \boldsymbol{ \chi(\theta)\sum_{\ell=1}^{L-1} (-1)^{\ell+1} \hat{S}_{\ell,\ell+1}^z }
    \end{align}\\ \vspace{-16pt}
    \begin{flushleft}   
    \hangafter 1
    \hangindent 1em
    \noindent 
    $\ell$ \& $L$: index \& number of matter sites,\\
    $\hat{\psi}_\ell$: fermionic matter field on site $\ell$ with mass $m$, \\
    $\hat{S}^+_{\ell,\ell+1}$: spin-1/2 operator on the link between sites $\ell$ and $(\ell+1)$, \\
    $\hat{E}_{\ell,\ell+1}=(-1)^{\ell+1}\hat{S}^z_{\ell,\ell+1}$: electric flux on the link, \\
    $\kappa$: matter--gauge coupling strength, \\ \vspace{1pt} 
    $\boldsymbol{ \chi(\theta)=g^2(\theta-\pi)/(2\pi)}$: \textbf{topological $\theta-$term} with $g$ the gauge coupling factor. 
    \end{flushleft}}  
    &
    \parbox[h]{0.31\textwidth}{\begin{flushleft}
    \hangafter 1
    \hangindent 1em
    \noindent
   When $U\approx\Delta\gg J$, among $\{$$\ket{20}$, $\ket{11}$, $\ket{12}$, $\ket{02}$, $\ket{01}$$\}$, only $|20\rangle$ and $|11\rangle$ are on resonance \\
    1. Mapping between the basis
    \vspace{-7pt}
   \begin{displaymath}\begin{array}{l|l} \hat{S}_{\ell,\ell+1}^z & \text{Fock states}\\ \hline
   \ket{\frac{1}{2}} &\ket{20} \\
   \ket{-\frac{1}{2}} & \ket{11}, \ket{12}, \ket{02}, \text{or} \ket{01}   \end{array} \end{displaymath} \\ \vspace{-2pt}
    2. U(1) gauge symmetry emerges, Gauss's law 
    $\hat{G}_\ell=(-1)^\ell\big(\hat{\psi}^\dagger_\ell\hat{\psi}_\ell+\hat{S}^z_{\ell-1,\ell}+\hat{S}^z_{\ell,\ell+1}\big)$ formulated from 
    $\big[\hat{H},\hat{G}_\ell\big]=0,\,\forall\ell$; 
    Here $\hat{G}_\ell$: $\hat{G}_\ell\ket{\phi}=g_\ell\ket{\phi},\,\forall \ell$ and 
    $g_\ell$ are background charges of local gauge sectors, \\
    3. for \textit{physical sector}: $g_\ell=0,\,\forall\ell$, leads to $\hat{\psi}^\dagger_\ell\hat{\psi}_\ell = -\hat{S}^z_{\ell-1,\ell}-\hat{S}^z_{\ell,\ell+1}$, \\ \vspace{4pt} 
    4. $\kappa \approx \sqrt{2}J,~m = (\Delta-U)/2,~\boldsymbol{\chi = \delta}$.
    \end{flushleft}}  
    &
    \parbox[h]{0.30\textwidth}{\begin{align}\nonumber
    \hat{H}_{\mathrm{BHM}} =&  -J \sum_{\ell=1}^{L-1} \left( \hat{b}_\ell^{\dagger} \hat{b}_{\ell+1} + \text{H.c.} \right) \\ \nonumber
    & + \sum_{\ell=1}^L \left[ \frac{U}{2}\hat{n}_\ell(\hat{n}_\ell-1) + \ell \Delta \hat{n}_\ell \right] \\ \nonumber%
    & + \boldsymbol{ \sum_{\ell=1}^L (-1)^\ell \frac{\delta}{2} \hat{n}_\ell }
    \end{align}
    \begin{flushleft}   
    \hangafter 1
    \hangindent 1em
    \noindent
    $\ell$ \& $L$: index \& number of bosonic sites,\\
    $\hat{b}_\ell^{(\dagger)}$: ladder bosonic operator on-site $\ell$, \\
    $\hat{n}_\ell = \hat{b}_\ell^{\dagger} \hat{b}_\ell$: number operator, \\
    $U$: on-site interaction strength,\\
    $J$: hopping amplitude,  \\ 
    $\delta$: staggered superlattice potential, \\ 
    $\Delta$: linearly tilted potential. \\
    \end{flushleft}
    } \\
    \hline
    \end{widetabular}
\end{table*}

Here, we report the implementation of a tunable topological $\theta$-angle in a $\mathrm{U}(1)$ gauge theory on an atom-number-resolved optical-lattice quantum simulator, through which we directly observe the confinement--deconfinement transition. Our experiment integrates a spin-dependent superlattice with the pattern-programmable trap potentials, allowing parallel and localized manipulation of atoms. This is crucial to realize the mapping from the interacting bosonic system in the presence of a staggered superlattice onto the target $\mathrm{U}(1)$ gauge theory possessing uniquely a dynamical electric field and thereafter leading to a tunable topological $\theta$-angle term; cf.~Fig.~\ref{figure1:models}a. We generated particle--antiparticle pairs with site-resolved addressing techniques and then monitored the \textit{in situ} propagating dynamics of these pairs. We observed constrained (ballistic) propagation for finite (vanishing) deviation of the $\theta$-angle from $\pi$, the microscopic smoking gun for the (de)confined phase. Moreover, we found that the confinement--deconfinement transition accompanies the previously observed Coleman's phase transition \cite{Yang2020observation} along $\theta = \pi$ upon tuning the fermionic rest mass.

\textbf{\textit{Models.---}}The target $\mathrm{U}(1)$ gauge theory in this work is a \textit{quantum link model} (QLM) \cite{Chandrasekharan1997} formulation of $1+1$D QED with a topological $\theta$-angle on a lattice \cite{Halimeh2022tuning,Cheng2022tunable}. To simulate this model, especially realizing the topological $\theta$-angle, we implement three major experimental ingredients: a Bose--Hubbard model (BHM), a linearly tilted potential \cite{Su2022observation} and a staggered potential with a superlattice to manipulate spinless ultracold bosons ($^{87}$Rb) as shown in Fig.~\ref{figure1:models} and Table \ref{tab:mapping}, where the Fock state of every neighboring sites $\ell$ and $\ell+1$ in the BHM maps to spin configuration in the QLM. Based on Gauss's law, in the \textit{physical sector} $g_\ell=0$ for all sites $\ell$, the electric charge at each site is then obtained as $\hat{Q}_\ell=(-1)^\ell\hat{\psi}_\ell^\dagger \hat{\psi}_\ell= \hat{E}_{\ell,\ell+1}-\hat{E}_{\ell-1,\ell}$, reducing experimental overhead as the matter sites and gauge links are simultaneously encoded in a correlated manner. As it becomes apparent from the mapping in Table \ref{tab:mapping}, we can realize a tunable topological $\theta$-angle in the experiment through a programmable staggered potential $\delta$.
 
The phase diagram of the $\mathrm{U}(1)$ QLM in the physical sector is shown in Fig.~\ref{figure1:models}b. At $\chi=0$, it hosts a global $\mathbb{Z}_2$ translation symmetry related to charge and parity conservation~\cite{Coleman1976} (but the latter is broken at finite $\chi$) and maps onto the PXP model under the constraints $g_\ell=0$ \cite{Surace2020lattice}.  
Although a special case of confinement--deconfinement transition occurs by tuning the mass along $\theta = \pi$ (i.e., $\chi  = 0$), when $\theta$ is tuning away from $\pi$ (i.e., $\chi\neq0$), the general case of confinement arises \cite{Halimeh2022tuning,Cheng2022tunable}, which can be accompanied by various nonergodic regimes such as Hilbert-space fragmentation and novel quantum scarring \cite{Desaules2023ergodicity}.

\begin{figure*}[t!]
    \centering     %
    \includegraphics[width=0.992\textwidth]{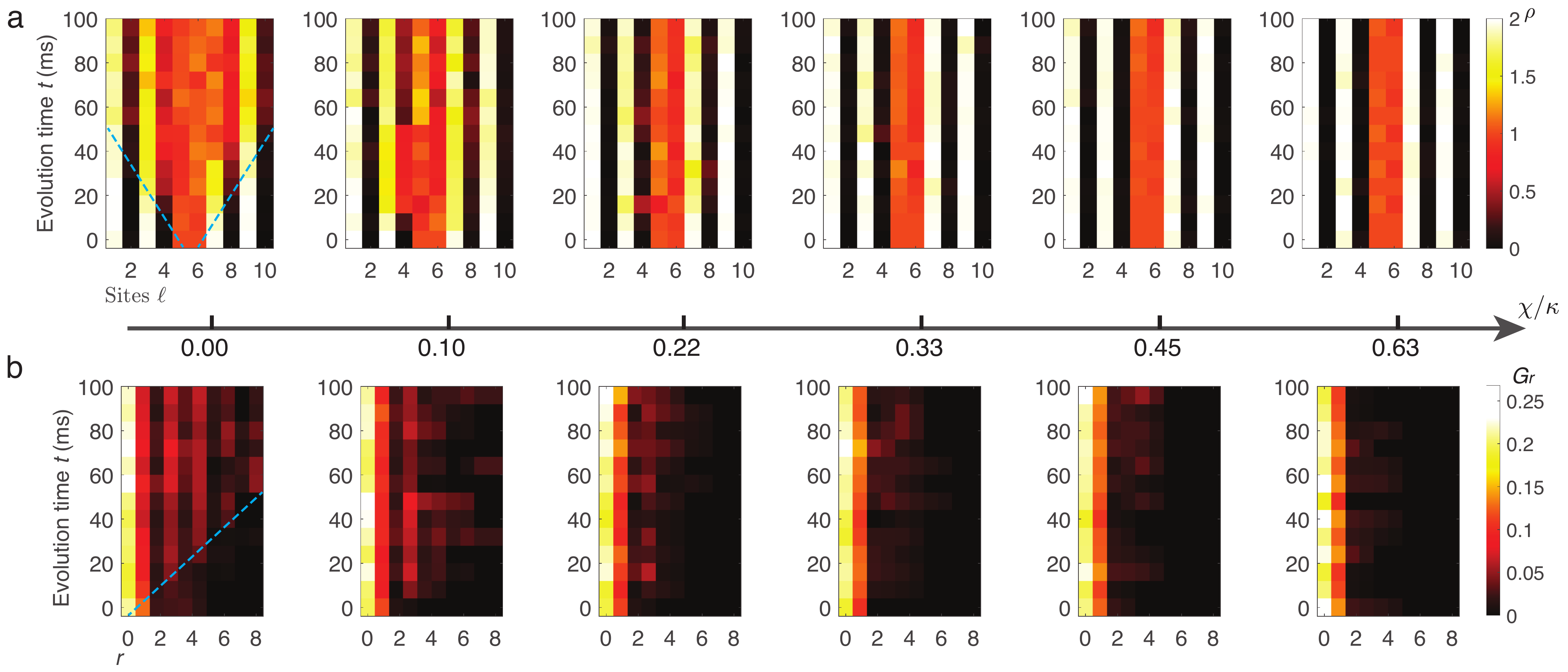}
    \caption{\textbf{Evolution of densities and two-point correlations along a tunable topological $\theta$-angle, specified by its deviation $\chi$ from 0.}
    (a) While at $\chi=0$ the atomic density profile indicates ballistic propagation of the particle--antiparticle pair, as $\chi/\kappa$ is increased, the evolution of the atomic density profile is significantly suppressed, indicating a bound pair.
    The blue dash lines in the figure are a guide for the eye. 
    (b) The same behavior can be observed in the two-point correlation function, which goes from a ballistic spread in time at $\chi=0$ to a localized profile at larger values of $\chi/\kappa$. 
    }
    \label{figure2:densities}
\end{figure*}

\textbf{\textit{Observation of confinement via microscopic dynamics.---}}
Our experimental realization uses the apparatus described previously in Ref \cite{Zhang2022quantum}, with which we prepare $^{87}$Rb atoms into defect-free Mott-insulator arrays with a filling factor of 99.2\%. 
The atoms are then initialized into a Fock state of the Hamiltonian $\hat{H}_{\mathrm{BHM}}$, corresponding to creating a particle--antiparticle pair on top of the ground state of Hamiltonian $\hat{H}_{\mathrm{QLM}}$ based on the mapping sketched in Fig.~\ref{figure1:models}a. 
Subsequently, we monitor the microscopic dynamics of particle--antiparticle separation behavior \cite{Halimeh2022tuning,Cheng2022tunable}, utilizing multiple observables to distinguish between confinement and deconfinement features. 
The ensuing analysis first centers on the microscopic dynamics of particle--particle pairs as the topological $\theta$-angle varies.

\textit{Confinement through tuning topological $\theta$-angle.---} 
We initially prepare the bosonic Fock state deterministically as $\ket{2020112020}$ with pattern-programmable addressing beams (see Methods), under the mapping sketched in Fig.~\ref{figure1:models}a.
This process corresponds to generating a particle--antiparticle pair on top of the vacuum state, which constitutes the ground state of Hamiltonian $\hat{H}_{\mathrm{QLM}}$ at $m/\kappa\to\infty$. 
Numerical results show that this configuration considerably overlaps with, and is thus an excellent approximation to, the target low-energy excited state when $m/\kappa\gtrsim 2$ (see Methods). 
In each experiment run, we prepare $\sim 5$ copies of identical 1D chains along the $y$ direction, each consisting of 10 atoms and accommodating 9 gauge field sites.
The tunable topological $\theta$-angle is implemented by applying staggered potential $\chi(\theta)=\delta$ with variable amplitudes. 
Subsequently, we quench the system to a fixed parameter region with $m \approx 4 \kappa$ within $0.5$ ms, where $\kappa \approx 50(1)$ Hz, and let the dynamics evolve to various final times. 
After that, we suddenly freeze the system and perform the single-site-resolved detection to obtain the atom number at each site \cite{Zhang2022quantum,Wang2022interrelated}. 
To mitigate the undesired effects from processes beyond the physical sector, we post-select our data with the rules \cite{Mildenberger2022,Wang2022interrelated}: (i) conservation of total atom number, and (ii) conservation of Gauss's law at all local constraints.

\begin{figure*}[t!]
    \centering     %
    \includegraphics[width=0.96\textwidth]{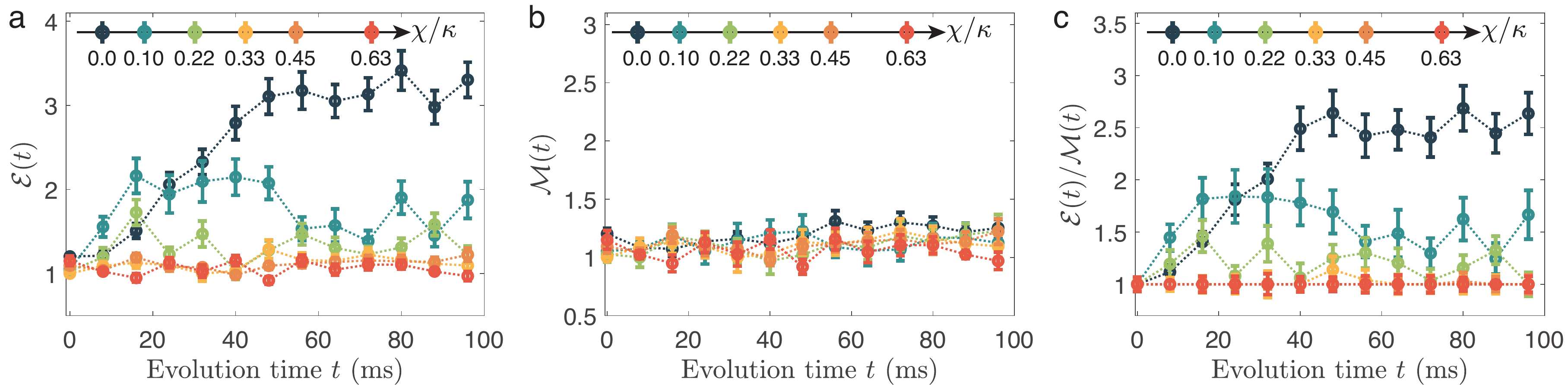}
    \caption{\textbf{Microscopic dynamics of the particle--antiparticle pair.} 
    (a) The extracted time-resolved results of the deviation $\mathcal{E}(t)$ in the electric field with respect to the vacuum state. Whereas $\mathcal{E}(t)$ grows roughly linearly in time before the particle--antiparticle pair reaches the chain boundaries for $\chi = 0$, its growth is gradually suppressed as $\chi/\kappa$ is increased. 
    (b) On the other hand, the deviation $\mathcal{M}(t)$ in the total charge relative to the vacuum state fluctuates roughly around unity regardless of the value of $\chi/\kappa$.
    (c) The extracted time-resolved results for $\mathcal{E}(t)/\mathcal{M}(t)$ at various values of $\chi/\kappa$ indicate the same behavior as $\mathcal{E}(t)$. 
    Error bars denote the standard error of mean (s.e.m.) in (a-c).
    }
    \label{figure3:observables}
\end{figure*}

The measured site-resolved atomic density $\langle\hat{n}_\ell(t)\rangle$ in the bosonic model $\hat{H}_{\mathrm{BHM}}$ is directly associated with the local flux (or, equivalently, the charge) configuration of the time-evolved wave function within the QLM picture. As illustrated in Fig.~\ref{figure2:densities}a, one can read out that the particle--antiparticle pair spreads ballistically over the entire lattice when $\chi = 0$. However, as $\chi/\kappa$ is increased, the evolution of the atomic density profile is significantly constrained, with the particle and antiparticle seemingly bound together throughout all investigated evolution times. This can be understood by noting that the topological $\theta$-angle term in Hamiltonian $\hat{H}_{\mathrm{QLM}}$ serves as an energy penalty for the growth of the electric string between the particle and antiparticle. At $\chi=0$, this energy penalty is absent, and thus the particle and antiparticle are unbound and can ballistically spread far from each other. In contrast, as $\chi$ is tuned to larger values, it is energetically unfavorable for the particle and antiparticle to spread away from each other, as then Gauss's law necessitates that the electric string between them grows; see Fig.~\ref{figure1:models}b. 
At the same time, we extract the QLM equal-time two-point charge correlation, which reads as
\begin{align}
    \label{eq:correlation}
    G_r(t) = \sum_\ell \big\langle\big[\hat{Q}_\ell(t) - \bar{Q}^\text{gs}_\ell\big]\big[\hat{Q}_{\ell+r}(t)-\bar{Q}^\text{gs}_{\ell+r}\big]\big\rangle,
\end{align}
where $\bar{Q}^\text{gs}_\ell$ is the mean charge at matter site $\ell$ in the ground state (i.e., before introducing the particle--antiparticle excitation). Fig.~\ref{figure2:densities}b shows the measured $G_r(t)$ for various values of $\chi/\kappa$. Similar to the density distribution, the spreading of the two-point correlation is also highly suppressed with increasing $\chi/\kappa$.

Moreover, we further analyze the time-evolved \textit{deviation} of the electric flux and charge from their values in the ground state using 
\begin{subequations} 
\label{eq:deviations}
\begin{align}
\mathcal{E}(t) & = \sum_{\ell=1}^{L-1}\big\langle\big[\hat{E}_{\ell,\ell+1}(t) - \bar{E}^\text{gs}_{\ell,\ell+1} \big]^2\big\rangle, \\ 
\mathcal{M}(t) & = \frac{1}{2} \sum_{\ell=1}^{L} \big\langle\big[\hat{Q}_\ell(t) - \bar{Q}^\text{gs}_\ell \big]^2\big\rangle,
\end{align}
\end{subequations}
where $\bar{E}^\text{gs}_{\ell,\ell+1}$ represents the mean electric flux at the link between sites $\ell$ and $\ell+1$ in the ground state. The first observable $\mathcal{E}(t)$ can be considered a measure of the length of the electric string between the particle and antiparticle, which at $t=0$ is $1$. It can grow to the system size in the deconfined phase, whereas it will saturate at a limited value in the confined phase. The second observable $\mathcal{M}(t)=G_0(t)/2$  counts the number of particle--antiparticle pairs in this case, which, due to the large value of $m/\kappa$, should remain roughly at its initial value throughout the dynamics regardless of the value of $\chi/\kappa$.

We plot $\mathcal{E}(t)$ and $\mathcal{M}(t)$ as a function of evolution time at various values of $\chi/\kappa$ in Fig.~\ref{figure3:observables}a,b.
We can see that when $\chi = 0$, which corresponds to $\theta = \pi$, $\mathcal{E}(t)$ increases roughly linearly at early times before approaching a plateau around $t\approx55$ ms when the spreading pair reaches the edges of the chain.
However, as we tune $\chi$ to finite values, we can see that the increase in $\mathcal{E}(t)$ is suppressed, and when $\chi/\kappa\gtrsim0.3$, $\mathcal{E}(t)$ shows little growth from its initial value. On the other hand, $\mathcal{M}(t)$ always fluctuates around unity for various $\chi/\kappa$, indicating that few additional particle--antiparticle pairs are created or annihilated during the time evolution due to the large-mass quench. 
Nevertheless, to remove the effect of unwanted creation and annihilation of particle--antiparticle pairs on $\mathcal{E}(t)$, we also plot $\mathcal{E}(t)/\mathcal{M}(t)$ in Fig.~\ref{figure3:observables}c. This ratio can be thought of as the length of the electric string per particle--antiparticle pair, and it clearly shows that the initial particle--antiparticle pair is unbound and can ballistically spread when $\chi=0$, while it is bound at finite values of $\chi$. These results coincide with the phase diagram illustrated in Fig.~\ref{figure1:models}b, and demonstrate the confinement--deconfinement transition through tuning the topological $\theta$-angle.

\begin{figure}[t!]
    \centering     %
    \includegraphics[width=0.48\textwidth]{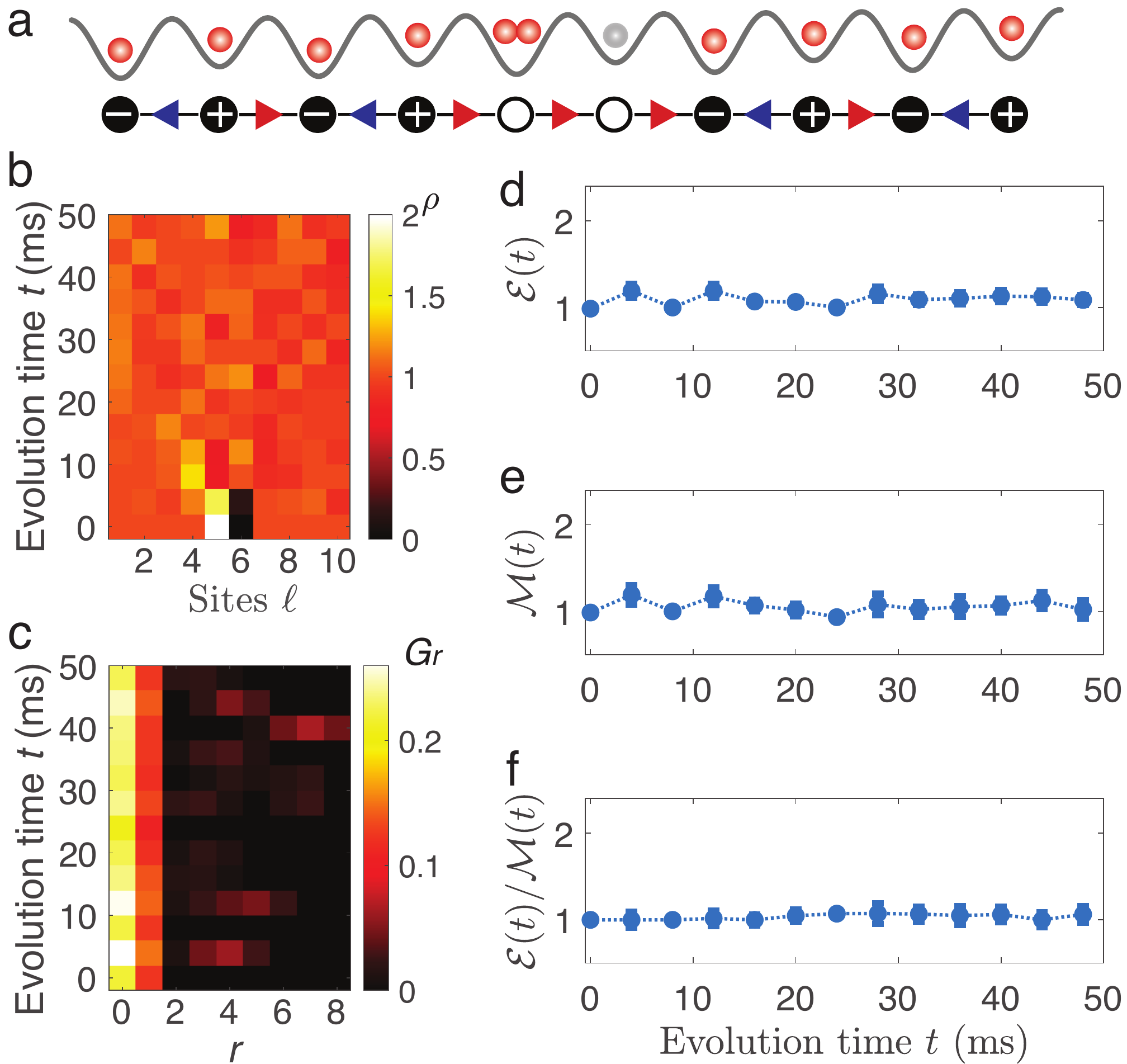}
    \caption{\textbf{Microscopic dynamics of the hole pair.} (a) Schematic illustration of the initial state, a hole--pair excitation on top of the charge-proliferated state. (b) The time-resolved density profile of the quench dynamics within the negative-mass regime quickly becomes homogeneous, indicating that the hole pair, though bound, can still propagate all over the lattice.
    (c) The corresponding two-point correlation function shows that the hole pair, though mobile, is bound.
    The extracted time-resolved results of (d) $\mathcal{E}(t)$, (e) $\mathcal{M}(t)$, and  (f) $\mathcal{E}(t)/\mathcal{M}(t)$ confirm this picture.
    Error bars denote the s.e.m.~in (d-f).
    }
    \label{figure4:confs}
\end{figure}

\textit{Confinement in the negative mass region}.---We now turn our attention to the negative-mass regime, where we start in the charge-proliferated state of the QLM and add a low-energy excitation in the form of a hole pair to it within the physical sector $g_\ell=0$. Under the mapping sketched in Fig.~\ref{figure1:models}a, in our quantum simulator, this is equivalent to preparing the configuration $\ket{1111201111}$ (see Methods and Fig.~\ref{figure4:confs}a). 
This configuration also significantly overlaps the target low-energy excited state at $m/\kappa\lesssim -4$ (see Methods).
After that, we quench the system to a large negative mass regime with $m \approx -4\kappa$ and let the system evolve, where $\kappa \approx 45(1)$ Hz.
We keep $\chi = 0$, corresponding to $\theta = \pi$, throughout the evolution. 
Finally, we suddenly freeze the system and detect the atom number at each site.

The measured time-resolved density profile and two-point correlation function are plotted in Fig.~\ref{figure4:confs}b,c.
The average density profile quickly evolves into a homogeneous distribution, while the two-point correlation always maintains a localized form.
These results indicate that the hole--hole pair is free to propagate, but the constituent holes are always bound together, as evidenced by the correlation function being highly localized around small $r$. 
We also plot the observables $\mathcal{E}(t)$ and $\mathcal{M}(t)$ along with their ratio $\mathcal{E}(t)/\mathcal{M}(t)$ in Fig.~\ref{figure4:confs}d-f.
We can see that the electrical string length per hole-pair is always kept at constant values. It illustrates that the distance between the two holes does not change during the time of evolution. The above results are indicative of confined dynamics.

Even though in this case $\chi=0$, a large negative mass and Gauss's law conspire to give rise to confinement. Two adjacent holes can only separate if the sites between them become empty as well, due to Gauss's law. However, the large negative mass makes the annihilation of particle--antiparticle pairs energetically unfavorable, which in turn leads to confinement.

\textbf{\textit{Conclusion and outlook}}.---We have directly observed the confinement--deconfinement transition in a $\mathrm{U}(1)$ quantum-link gauge theory via monitoring the microscopic dynamics of particle--antiparticle pairs with tunable topological $\theta$-angle. Meanwhile, a confined phase has also been found without tuning the topological $\theta$-angle, where a large negative mass and Gauss's law lead to the constrained dynamics of the hole--pair excitations. These findings are based on our quantum simulator, combining a site-resolved quantum gas microscope with a pattern-programmable optical superlattice and possessing versatile capabilities to generate arbitrary initial state, flexibly tailoring trap potentials, and time-space-number-resolved detections. 

Our experimental observations are consistent with numerical benchmarks, demonstrating the validity of our quantum simulator quantitatively and therefore, it can serve as a powerful platform for future explorations of the rich physics of gauge theories, e.g., string breaking \cite{Surace2020lattice,Banerjee2012atomic}, false vacuum decay \cite{Lagnese2021false}, and quantum thermalization in gauge theories \cite{Zhou2022thermalization,Berges2021qcd}. Additionally, we can explore the influence of a tunable topological $\theta$-angle on condensed-matter phenomena, such as dynamical quantum phase transitions \cite{Zache2019dynamical,Huang2019dynamical}, disorder-free localization \cite{Smith2017,Brenes2018}, and other ergodicity-breaking phenomena \cite{Desaules2023ergodicity}. By further optimizing the trap potentials to mitigate the effects of inhomogeneities, we can increase the size of our quantum simulator several times, potentially achieving quantum computational advantage over classical simulations. Furthermore, our current scheme for implementing quantum link models can be extended to higher spin truncations \cite{Osborne2023SpinS} and higher spatial dimensions \cite{Osborne2022large}, leading to the observation of even richer exotic gauge-theory physics.

\textit{Acknowledgements}.---We thank Hui Zhai for discussions. This work was supported by NNSFC grant 12125409, Innovation Program for Quantum Science and Technology 2021ZD0302000. J.C.H.~acknowledges funding within the QuantERA II Programme that has received funding from the European Union’s Horizon 2020 research and innovation programme under Grand Agreement No 101017733, support by the QuantERA grant DYNAMITE, by the Deutsche Forschungsgemeinschaft (DFG, German Research Foundation) under project number 499183856, funding by the Deutsche Forschungsgemeinschaft (DFG, German Research Foundation) under Germany's Excellence Strategy -- EXC-2111 -- 390814868, and funding from the European Research Council (ERC) under the European Union’s Horizon 2020 research and innovation programm (Grant Agreement no 948141) — ERC Starting Grant SimUcQuam. Y.C.~acknowledges support from NSFC Grant No. 12204034, Fundamental Research Funds for the Central Universities (No.FRFTP-22-101A1).



    
\bibliography{lgt_conf.bib}    
\bibliographystyle{naturemag}   

\onecolumngrid
\vspace*{0.5cm}
\newpage
\begin{center}
    \textbf{METHODS AND SUPPLEMENTARY MATERIALS}
\end{center}
\vspace*{0.5cm}

\twocolumngrid
\tableofcontents
\appendix
\setcounter{secnumdepth}{2}

\twocolumngrid
\setcounter{equation}{0}
\setcounter{figure}{0}
\makeatletter
\makeatother
\renewcommand{\theequation}{S\arabic{equation}}
\renewcommand{\thefigure}{S\arabic{figure}}
\renewcommand{\thetable}{S\arabic{table}}

\section{Target $\mathrm{U}(1)$ gauge theory}

In this work, we focus on a $\mathrm{U}(1)$ lattice gauge theory (LGT), the lattice version of one-dimensional (1D) quantum electrodynamics (QED) \cite{Schwinger1951}. 
Such lattice Schwinger model can be described by the following Hamiltonian \cite{Kogut1979}
\begin{align}\nonumber
    \hat{H}_{\mathrm{LGT}} = & -\kappa \sum_{\ell=1}^{L-1} \left( \hat{\psi}_\ell^{\dagger} \hat{U}_{\ell,\ell+1} \hat{\psi}_{\ell+1} + \text{H.c.} \right) \\
    \label{eq:ham_lgt}
    & + m \sum_{\ell=1}^L (-1)^\ell \hat{\psi}_\ell^{\dagger} \hat{\psi}_\ell + \sum_{\ell=1}^{L-1} \left( \hat{E}_{\ell,\ell+1} + E_{\mathrm{bg}} \right)^2,
\end{align}
\noindent where $\ell$ labels the matter sites, and $\hat{\psi}_\ell^{\dagger}$, $\hat{\psi}_\ell$ are the creation and annihilation operators of the fermionic matter field on site $\ell$ of the 1D lattice.
$\hat{U}_{\ell,\ell+1}$ and $\hat{E}_{\ell,\ell+1}$ are the parallel transporter and electric field operators, respectively, representing a $\mathrm{U}(1)$ gauge field on the link connecting sites $\ell$ and $\ell+1$.  
They satisfy the commutation relations $[\hat{E}_{\ell,\ell+1},\hat{U}_{r,r+1}] = \delta_{\ell,r} \hat{U}_{\ell,\ell+1}$ and $[\hat{U}_{\ell,\ell+1},\hat{U}_{r,r+1}^\dagger]=0$. The homogeneous background field is $E_\text{bg}=g\theta/(2\pi)$, where $g$ is the gauge coupling and $\theta$ is the topological angle.
The above lattice Schwinger model (Eq.~\eqref{eq:ham_lgt}) hosts a $\mathrm{U}(1)$ gauge symmetry with generator
\begin{align}
    \label{eq:gauss_law}
    \hat{G}_\ell = \hat{E}_{\ell,\ell+1} - \hat{E}_{\ell-1,\ell} - g\left[ \hat{\psi}_\ell^{\dagger} \hat{\psi}_\ell - \frac{1-(-1)^\ell}{2} \right].
\end{align}

\subsection{Quantum link formulation of Schwinger model}

Next, we will consider the quantum link formulation of the above model and focus on the spin-$1/2$ representation \cite{Chandrasekharan1997}.
At this moment, we represent $\hat{U}_{\ell,\ell+1}$ and $\hat{E}_{\ell,\ell+1}$ by spin  variables, i.e., $\hat{U}_{\ell,\ell+1} = \hat{S}_{\ell,\ell+1}^{+}$ and $\hat{E}_{\ell,\ell+1} = g\hat{S}_{\ell,\ell+1}^{z}$.
Then, Hamiltonian~\eqref{eq:ham_lgt} can be rewritten as (up to an inconsequential constant) \cite{Halimeh2022tuning,Cheng2022tunable}
\begin{align}\nonumber
    \hat{H} = & -\kappa \sum_{\ell=1}^{L-1} \left( \hat{\psi}_\ell^\dagger \hat{S}_{\ell,\ell+1}^{+} \hat{\psi}_{\ell+1} + \text{H.c.} \right) \\
    \label{eq:qlm_sup}
    & + m \sum_{\ell=1}^L (-1)^\ell\hat{\psi}_\ell^{\dagger} \hat{\psi}_\ell + \chi \sum_{\ell=1}^{L-1}\hat{S}_{\ell,\ell+1}^z,
\end{align}
where $\chi=g^2(\theta-\pi)/(2\pi)$ denotes the deviation of the topological $\theta$-angle from $\pi$.
We can see that, $\chi = 0$ corresponds to $\theta = \pi$, and any finite $\chi$ tunes $\theta$ away from $\pi$.

Employing the particle--hole transformation \cite{Hauke2013}
\begin{subequations}
    \begin{align}
        &\hat{\psi}_\ell\to\frac{1+(-1)^\ell}{2}\hat{\psi}_\ell+\frac{1-(-1)^\ell}{2}\hat{\psi}_\ell^\dagger,\\
        &\hat{S}^{z(y)}_{\ell,\ell+1}\to(-1)^{\ell+1}\hat{S}^{z(y)}_{\ell,\ell+1},
    \end{align}
\end{subequations}
we arrive at the QLM Hamiltonian
\begin{align}\nonumber
    \hat{H} = & -\kappa \sum_{\ell=1}^{L-1} \left( \hat{\psi}_\ell \hat{S}_{\ell,\ell+1}^{+} \hat{\psi}_{\ell+1} + \text{H.c.} \right) \\
    & + m \sum_{\ell=1}^L \hat{\psi}_\ell^{\dagger} \hat{\psi}_\ell + \chi \sum_{\ell=1}^{L-1}(-1)^{\ell+1}\hat{S}_{\ell,\ell+1}^z,
\end{align}
outlined in Table~\ref{tab:mapping}.

\subsection{Mapping to Bose--Hubbard simulator}

In our experiment, the 1D Bose--Hubbard model with a linear tilt and a staggered potential, and under open boundary conditions, can be described by the Hamiltonian
\begin{align}\nonumber
    \hat{H}_{\mathrm{BHM}} = & -J \sum_{\ell=1}^{L-1} \left( \hat{b}_\ell^{\dagger} \hat{b}_{\ell+1} + \text{H.c.} \right) \\
    \label{eq:bhm_sup}
    & + \sum_{\ell=1}^L \left[ \frac{U}{2}\hat{n}_\ell(\hat{n}_\ell-1) + \epsilon_\ell \hat{n}_\ell \right],
\end{align}
where $\hat{b}_\ell^{\dagger}$, $\hat{b}_\ell$ are bosonic creation and annihilation operators, $\hat{n}_\ell = \hat{b}_\ell^{\dagger} \hat{b}_\ell$, $U$ is the on-site interaction, $J$ is the hopping amplitude, and $L$ is the number of lattice sites. 
The energy offset $\epsilon_\ell = (-1)^\ell \delta/2 + \ell \Delta$ consists of a staggered superlattice potential $\delta$, and a liner tilt potential $\Delta$.

When considering the near-resonant condition $U \approx \Delta \gg J$, and $\delta = 0$, the mainly allowed hopping process is $11 \leftrightarrow 20$. 
As a consequence, in the adjacent sites, only certain types of two-site configurations (20, 11, 12, 02, 01) are allowed, while all other two-site configurations (22, 21, 10, 00) are forbidden in the near resonance condition \cite{Su2022observation}.
Moreover, when we adopt configuration 20 as an excited state, labeled as $\bullet$, all other configurations are treated as the ground state, marked as $\circ$. 
The above-mentioned BHM (Eq.~\eqref{eq:bhm_sup}) maps to the following PXP model, read as \cite{Su2022observation}
\begin{equation}
    \label{eq:PXP}
    \hat{H}_{\mathrm{PXP}} = \sum_\ell \left[ -\kappa \hat{P}_{\ell-1,\ell} \hat{S}_{\ell,\ell+1}^x \hat{P}_{\ell+1,\ell+2} - 2m \hat{S}_{\ell,\ell+1}^z \right],
\end{equation}
\noindent where $\kappa \approx \sqrt{2}J$, $2m = \Delta-U$, and projectors $\hat{P}_{\ell,\ell+1} = \ketbra{\circ}$. 
Therefore, the unit-filling state $\ket{111111\ldots}$ maps to the polarized state $\ket{\circ \circ \circ \circ \circ \ldots}$, while another state $\ket{202020\ldots}$, corresponds to $\ket{\mathbb{Z}_2}=\ket{\bullet \circ \bullet \circ \bullet\ldots}$ state, which contains the maximum number of excitations constrained by the PXP model.

Theoretical works have already demonstrated that the above PXP model (Eq.~\eqref{eq:PXP}) is equivalent to the $\mathrm{U}(1)$ QLM (Eq.~\eqref{eq:qlm_sup}), taking into account that the gauge charges $\hat{G}_\ell=0$ for all sites $\ell$ \cite{Surace2020lattice}.
Therefore, the 1D-tilted Bose--Hubbard model (Eq.~\eqref{eq:bhm_sup}) realized in our experiment can be directly used for simulating the $\mathrm{U}(1)$ QLM (Eq.~\eqref{eq:qlm_sup}).
Moreover, we can realize a tunable topological $\theta$-angle in the experiment through a programmable staggered potential $\delta$.

\section{Experimental sequence}

\subsection{Mott insulator preparation}

Our experiment starts with the preparation of a two-dimensional (2D) Bose-Einstein condensate of $^{87}$Rb atoms in the $\ket{F=1,m_\mathrm{F}=-1}$ state, which has been discussed in our previous work \cite{Zhang2022quantum}. Subsequently, we utilize the staggered-immersion cooling method, as recently demonstrated, to prepare multiple copies of the one-dimensional (1D) near-unity filling Mott insulator state with a filling factor of 99.2(2)\% \cite{Zhang2022quantum}.

\begin{figure}[ht!]
    \centering     %
    \includegraphics[width=0.42\textwidth]{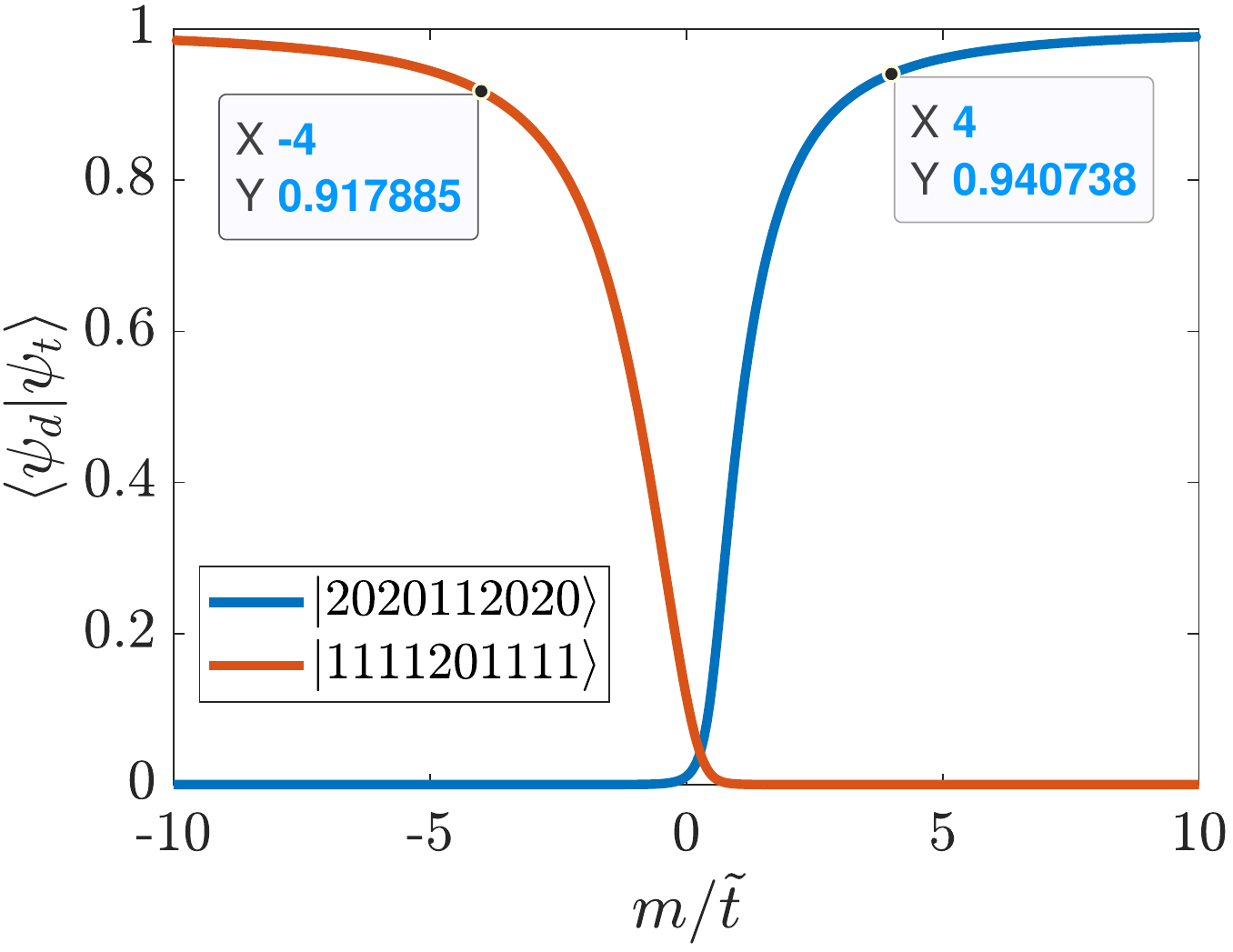}
    \caption{\textbf{Overlaps between the target states and the deterministic configurations}.
    The numerical overlaps between the low-energy excited state $\ket{\psi_t}$ and two deterministic states, namely $\ket{\psi_d}=\ket{2020112020}$ and $\ket{1111201111}$, are shown as solid blue and orange lines, respectively, for different values of $m/\kappa$.
    }
    \label{figureS1:overlaps}
\end{figure}

\begin{figure*}[ht!]
    \centering     %
    \includegraphics[width=0.74\textwidth]{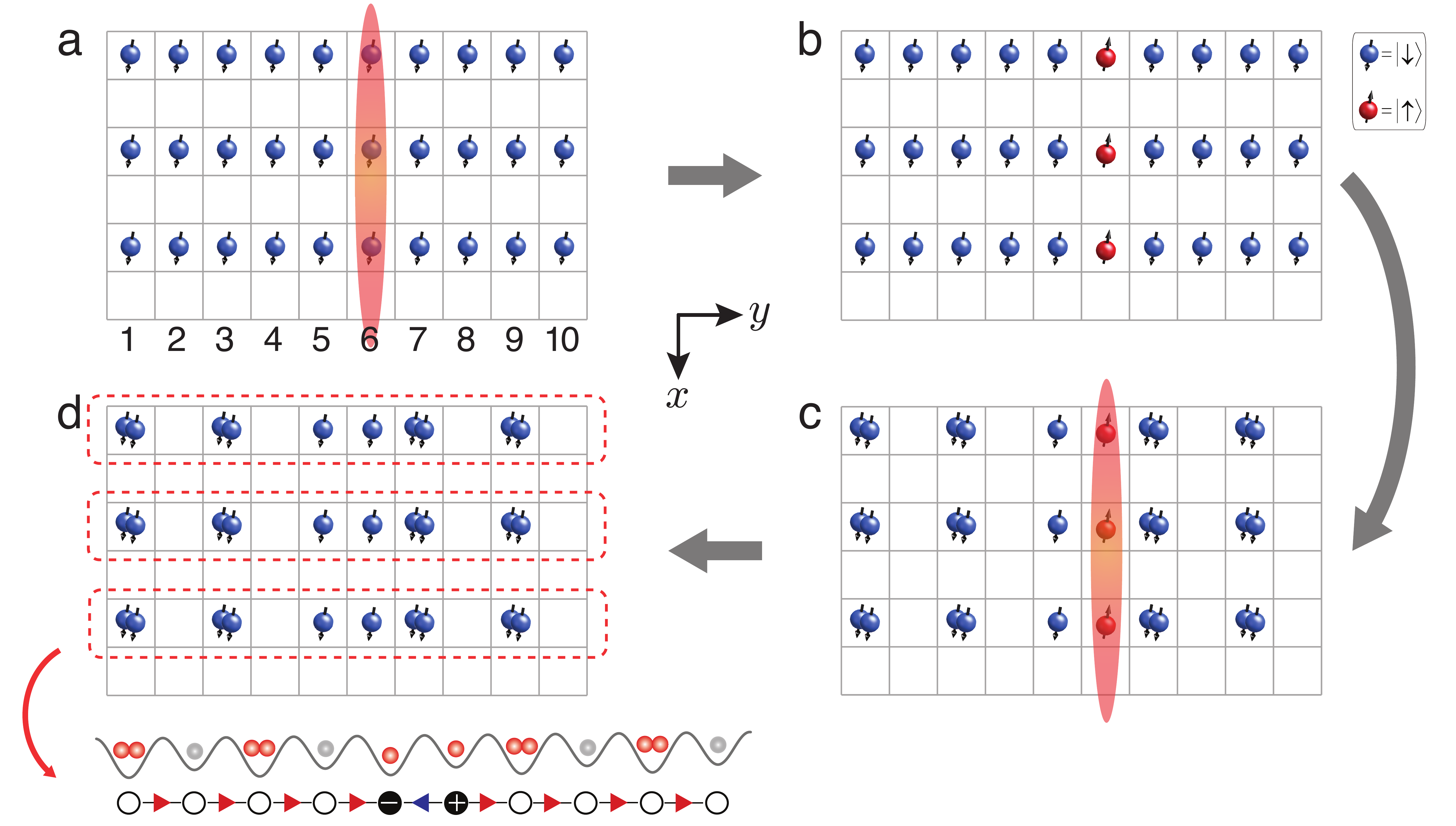}
    \caption{\textbf{Procedures of initial state preparation in the positive mass region}.
    (a) An illustration of the atom distribution of the prepared unity-filled Mott insulator state after staggered immersion cooling. Atoms in orange-shaded chains are then addressed by DMD light. 
    (b) An illustration of the atomic spin distribution after the site-dependent addressing. 
    (c) An illustration of the state after we merge every two atoms in $\ket{\downarrow}$ state onto a single site along the $y$ direction. 
    (d) The prepared copies of $\ket{2020112020}$ state along the $y$ direction.
    }
    \label{figureS2:init_stat_positive}
\end{figure*}

\subsection{Initial state preparation}

To observe the (de)confinement phenomenon of the ground states throughout the entire phase diagram of the $\mathrm{U}(1)$ QLM (as illustrated in Fig.~\ref{figure1:models}b), we need to prepare an appropriate target state. This can be achieved by introducing a low-energy excitation on top of the ground state to create or annihilate a particle--antiparticle pair ($\hat{S}_{l,l+1}^- \hat{b}_l^{\dagger} \hat{b}_{l+1}^{\dagger} \ket{\psi_\mathrm{g}}$ or $\hat{S}_{l,l+1}^+ \hat{b}_l \hat{b}_{l+1} \ket{\psi_\mathrm{g}}$) \cite{Halimeh2022tuning,Cheng2022tunable}. However, preparing a quantum many-body ground state in an experiment is generally challenging, let alone creating an excitation on it.

Fortunately, numerical calculations reveal that some deterministic configurations exhibit a significant overlap with the target states, as depicted in Fig.~\ref{figureS1:overlaps}. According to Fig.~\ref{figureS1:overlaps}, we observe that a configuration such as $\ket{\ldots201120\ldots}$ overlaps the low-energy excited state with fidelity over 90\% when $m/\kappa$ > 4, while another configuration, $\ket{\ldots112011\ldots}$, overlaps the low-energy excited state with fidelity over 90\% when $m/\kappa$ < -4.

\textbf{Positive mass region}. 
To begin our experiment, we start by preparing near-unity-filled Mott insulators (in the hyperfine state $\ket{\downarrow} = \ket{F=1,m_\mathrm{F}=-1}$). We then employ a spin-dependent superlattice along the $y$-direction and apply an addressing beam to a single well along the $x$-direction simultaneously \cite{Zhang2022quantum}.
The current state configuration is  $\ket{\ldots,\downarrow,\downarrow,\downarrow,\downarrow,\downarrow,\downarrow,\ldots}$, as illustrated in Fig.~\ref{figureS2:init_stat_positive}a.
Then, we apply two consecutive resonance microwaves to flip these atoms on the right side of the double wells from the $\ket{\downarrow}$ state to the $\ket{\uparrow}$ state and back to the $\ket{\downarrow}$ state, where $\ket{\uparrow} = \ket{F=2,m_\mathrm{F}=-2}$ \cite{Zhang2022quantum}.
Here, we turn off the addressing beam after applying the first microwave.
Now, the state configuration is $\ket{\ldots,\downarrow,\downarrow,\downarrow,\uparrow,\downarrow,\downarrow,\ldots}$, which is illustrated in Fig.~\ref{figureS2:init_stat_positive}b.
After that, we turn on the addressing beam again and merge the two atoms into the left side of the double well, leading to the current state configuration $\ket{\ldots,\downarrow\downarrow,0,\downarrow,\uparrow,\downarrow\downarrow,0,\ldots}$, as shown in Fig.~\ref{figureS2:init_stat_positive}c.
Finally, we turn off the addressing beam and flip the $\ket{\uparrow}$ state back to the $\ket{\downarrow}$ state using an additional microwave.
At this moment, we finish preparing the initial state configuration $\ket{\ldots201120\ldots}$, as depicted in Fig.~\ref{figureS2:init_stat_positive}d.

\begin{figure*}[ht!]
    \centering     %
    \includegraphics[width=0.74\textwidth]{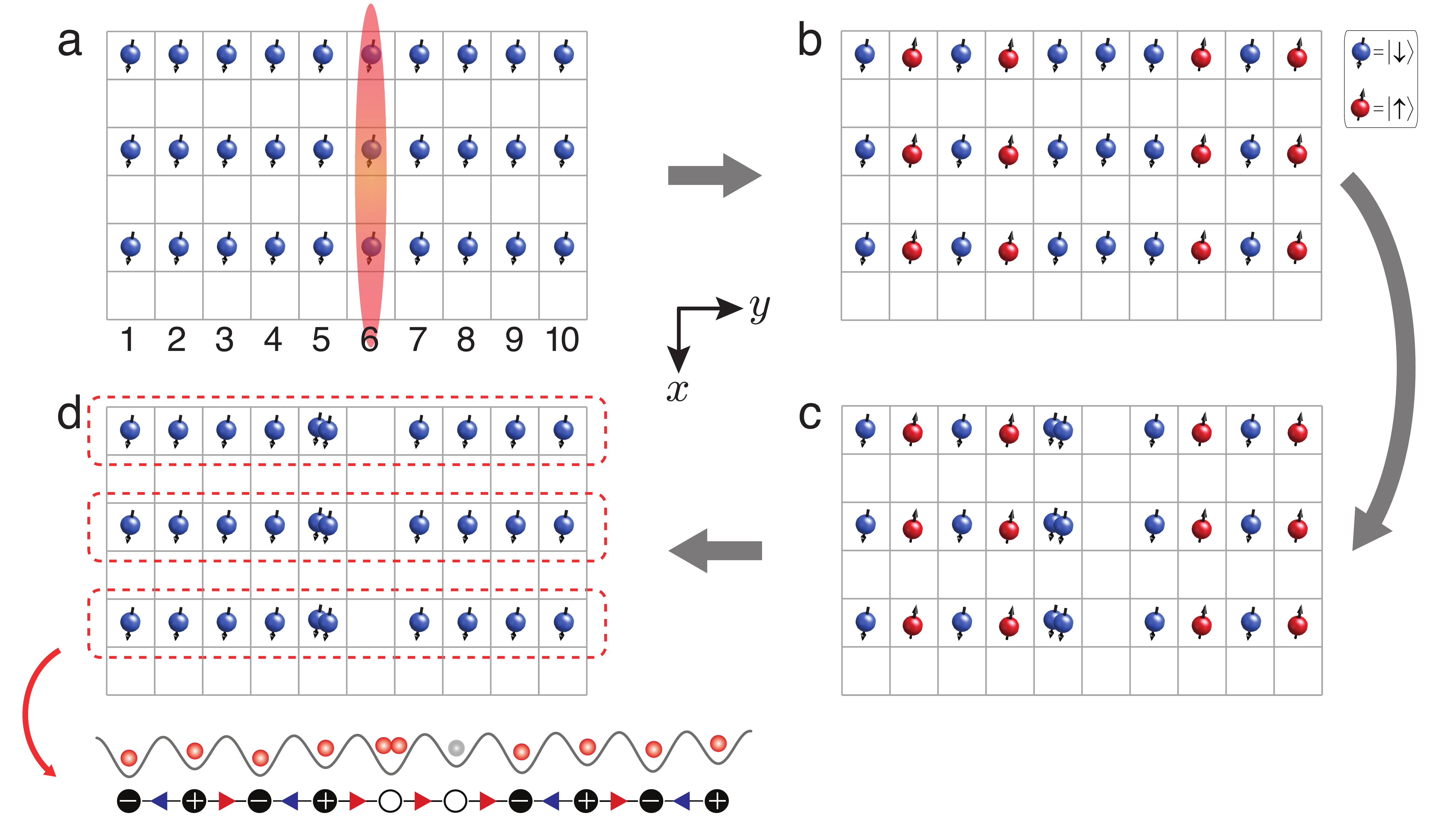}
    \caption{\textbf{Procedures of initial state preparation in the negative mass region}.
    (a) An illustration of the atom distribution of the prepared unity-filled Mott insulator state after staggered immersion cooling. Atoms in orange-shaded chains are then addressed by DMD light. 
    (b) An illustration of the atomic spin distribution after the site-dependent addressing. 
    (c) An illustration of the state after we adiabatically transfer every two atoms in $\ket{\downarrow}$ state onto a single site along the $y$ direction. 
    (d) The prepared copies of $\ket{1111201111}$ state along the $y$ direction.
    }
    \label{figureS3:init_stat_negative}
\end{figure*}

\textbf{Negative mass region}. 
After preparing near-unity-filled Mott insulators, we also employ a spin-dependent superlattice along the $y$-direction and apply an addressing beam to a single well along the $x$-direction simultaneously.
The current state configuration is  $\ket{\ldots,\downarrow,\downarrow,\downarrow,\downarrow,\downarrow,\downarrow,\ldots}$, as shown in Fig.~\ref{figureS3:init_stat_negative}a.
Then, we apply a single resonance microwave to flip these atoms on the right side of the double wells from the $\ket{\downarrow}$ state to the $\ket{\uparrow}$ state.
Now, the state configuration is $\ket{\ldots,\downarrow,\uparrow,\downarrow,\downarrow,\downarrow,\uparrow,\ldots}$, as illustrated in Fig.~\ref{figureS3:init_stat_negative}b.
After that, we only merge these two atoms in $\ket{\downarrow}$ state into the left side of the double-well (the details can be found in the following section \ref{section:adi_passage}).
The current state configuration is $\ket{\ldots,\downarrow,\uparrow,\downarrow\downarrow,0,\downarrow,\uparrow,\ldots}$, as shown in Fig.~\ref{figureS3:init_stat_negative}c.
Finally, we flip the $\ket{\uparrow}$ state back to the $\ket{\downarrow}$ state with an additional microwave.
At this moment, we finish preparing the initial state configuration $\ket{\ldots112011\ldots}$, as depicted in Fig.~\ref{figureS3:init_stat_negative}d.

\begin{figure*}[ht!]
    \centering     %
    \includegraphics[width=0.99\textwidth]{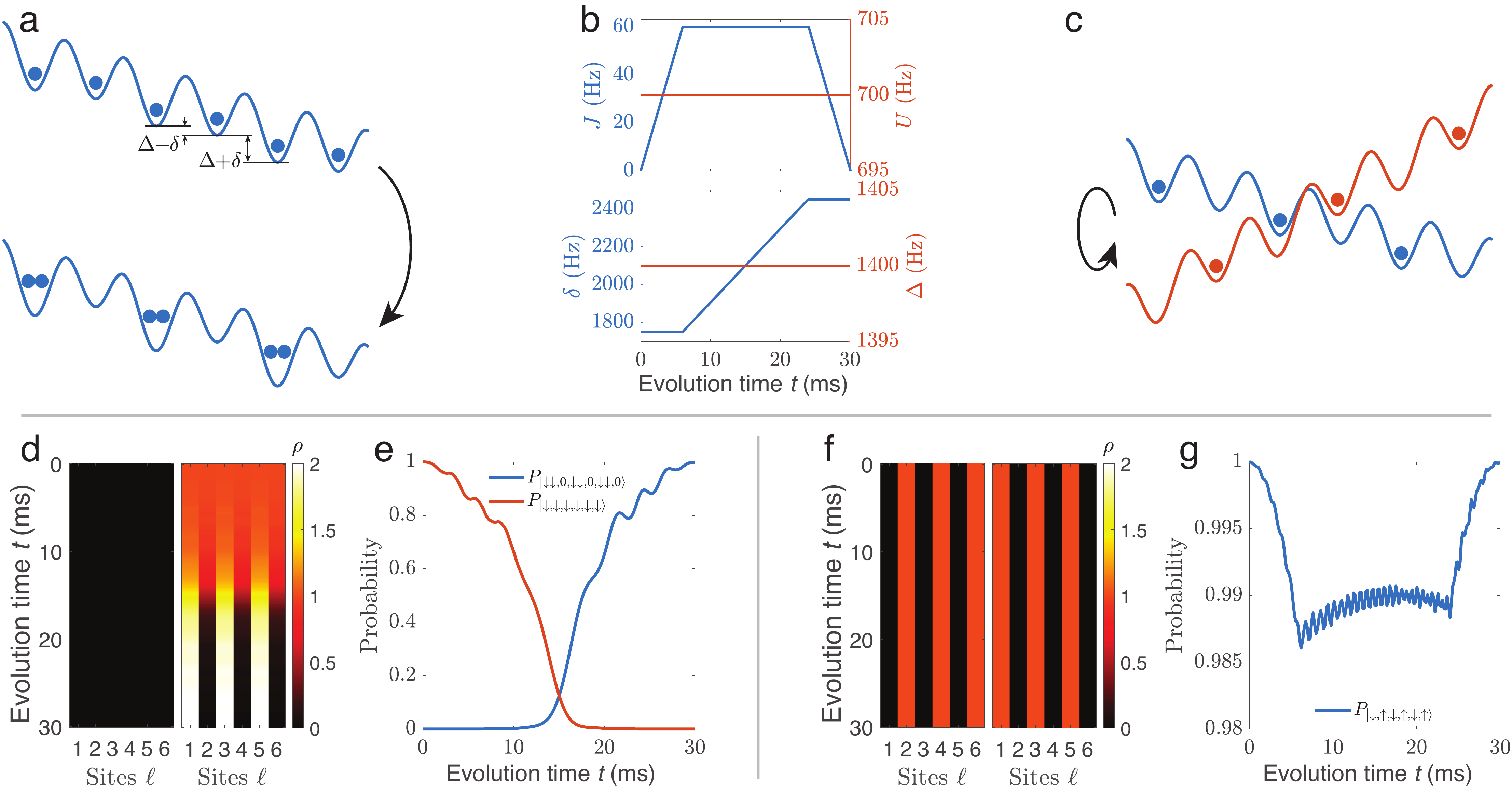}
    \caption{\textbf{Adiabatic passage}.
    (a) Schematic total optical lattice potentials for the $\ket{\downarrow}$ (solid blue line) atoms together with the relevant state-independent staggered potential $\delta$ and state-dependent tilt gradient $\Delta$ before and after the adiabatic passage process. Solid blue circles indicate an exemplary initial state of $\ket{\downarrow}$ atoms in the optical lattice. 
    The initial atom distribution $\ket{\downarrow,\downarrow,\downarrow,\downarrow,\downarrow,\downarrow}$ is transferred into $\ket{\downarrow\downarrow,0,\downarrow\downarrow,0,\downarrow\downarrow,0}$. 
    (b) The corresponding ramping protocols are employed in the adiabatic passage process. 
    (c) Schematic total optical lattice potentials for the $\ket{\downarrow}$ (solid blue line) and $\ket{\uparrow}$ atoms (solid orange line) together with the relevant state-independent staggered potential $\delta$ and state-dependent tilt gradient ($\Delta$ for $\ket{\downarrow}$ atoms, and $-2\Delta$ for $\ket{\uparrow}$ atoms) before and after the adiabatic passage process. Solid circles indicate an exemplary initial state of $\ket{\downarrow}$ (blue circles) and $\ket{\uparrow}$ atoms (orange circles) in the optical lattice. 
    The initial atom distribution $\ket{\downarrow,\uparrow,\downarrow,\uparrow,\downarrow,\uparrow}$ is kept unchanged. 
    (d) Numerical results of the time-resolved density profiles for $\ket{\uparrow}$ state (left) and $\ket{\downarrow}$ state (right), respectively. 
    (e) Probabilities of the atom distributions of $\ket{\downarrow,\downarrow,\downarrow,\downarrow,\downarrow,\downarrow}$ and $\ket{\downarrow\downarrow,0,\downarrow\downarrow,0,\downarrow\downarrow,0}$ during the adiabatic passage process. 
    (f) Numerical results of the time-resolved density profiles for $\ket{\uparrow}$ state (left) and $\ket{\downarrow}$ state (right), respectively. 
    (g) Probabilities of the atom distribution of $\ket{\downarrow,\uparrow,\downarrow,\uparrow,\downarrow,\uparrow}$ during the adiabatic passage process. 
    }
    \label{figureS9:adiabatic_passage}
\end{figure*}

\subsection{Quench and evolution}

After preparing the deterministic initial state configurations, we quench the system to the corresponding parameter conditions within 0.5 ms and then allow the system to evolve freely.

\subsection{Quantum state read out} 

To obtain the final quantum state, we first use the superlattice along the $x$-direction to split two atoms into two adjacent lattice sites of a double well, followed by a site-resolved atom number detection \cite{Zhang2022quantum}. 
To avoid parity projection during fluorescence imaging, we perform such an expansion process before the imaging procedure.
This allows us to resolve atom numbers up to two, starting from zero at a single site.
We then employ the same postselection rules as in our previous work \cite{Mildenberger2022,Wang2022interrelated}: (i) The total atom number remains the same as that of the initial state, and (ii) Gauss's law must be obeyed for all sites.

\begin{figure*}[ht!]
    \centering     %
    \includegraphics[width=0.96\textwidth]{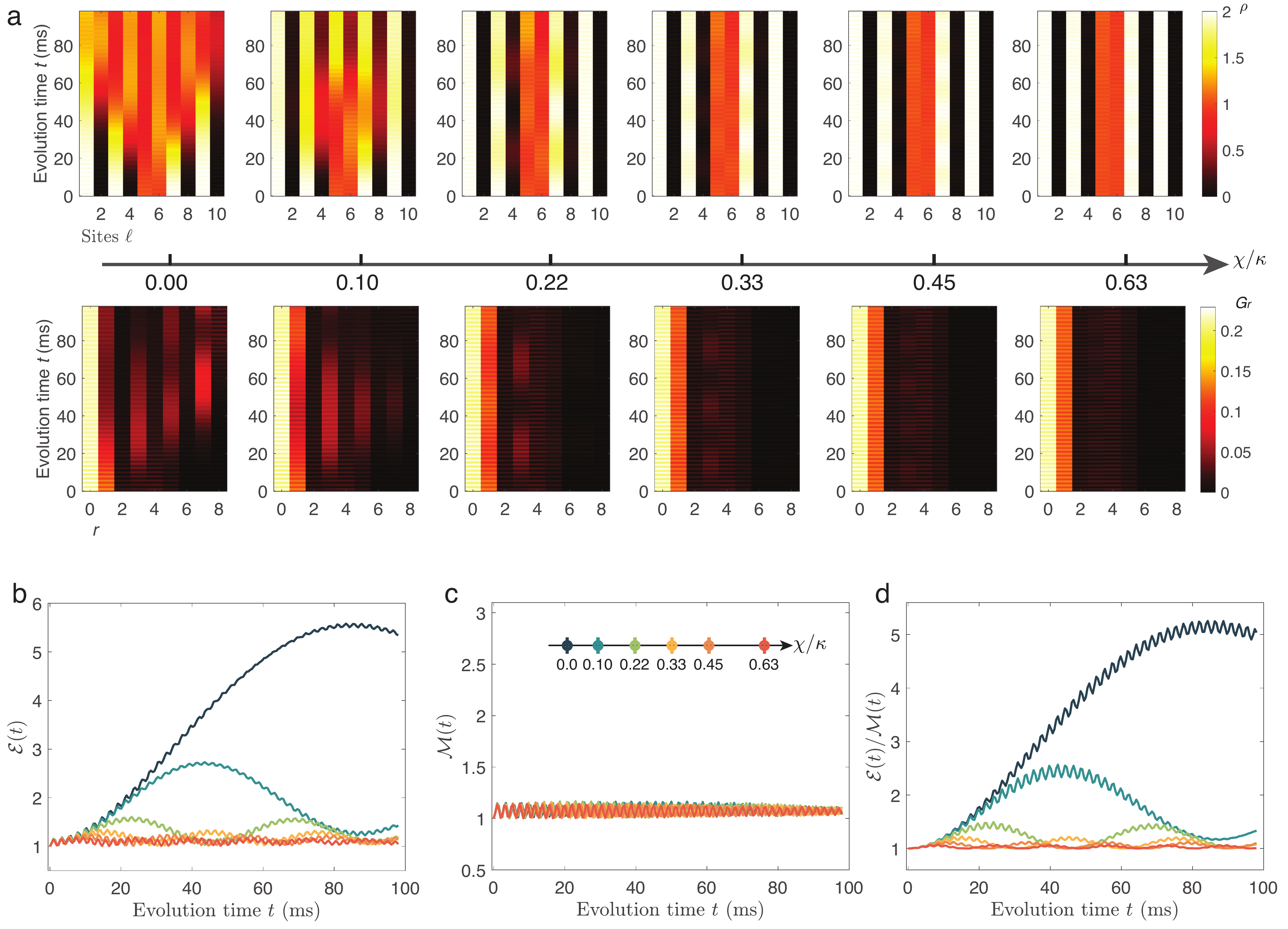}
    \caption{\textbf{Numerical results in the positive mass region}.
    (a) The time-resolved density profiles and two-point correlations at various $\chi/\kappa$.
    The extracted time-resolved results of the observables $\mathcal{E}(t)$ (b), $\mathcal{M}(t)$ (c) and $2\mathcal{E}(t)/\mathcal{M}(t)$ (d) at various $\chi/\kappa$. 
    }
    \label{figureS4:numerical1}
\end{figure*}

\section{Adiabatic passage for state preparation} \label{section:adi_passage}

In this section, we describe the method for adiabatically transferring a configuration with one $\ket{\downarrow}$ atom on each side of the double well to a configuration with both atoms on the left side of the double well, while keeping another configuration with one $\ket{\downarrow}$ atom occupied on the left side and one $\ket{\uparrow}$ atom on the right side of the double well unchanged, as illustrated in Fig.~\ref{figureS3:init_stat_negative}(b,c) and Fig.~\ref{figureS9:adiabatic_passage}(a,c), respectively.

\textbf{Procedures and numerical simulations}. 
As illustrated in Fig.~\ref{figureS9:adiabatic_passage}a, we first employed a state-independent staggered potential $\delta$ and a state-dependent linear tilt gradient $\Delta$ along the $y$-direction for the initial state distribution $\ket{\downarrow,\downarrow,\downarrow,\downarrow,\downarrow,\downarrow}$.
Then, we perform the adiabatic passage by ramping the tunneling strength $J$ and staggered potential $\delta$ while keeping the on-site interaction $U$ and linear tilt $\Delta$ constant, as shown in Fig.~\ref{figureS9:adiabatic_passage}(b). 
We linearly change the staggered potential $\delta$ from $U/2+\Delta$ to $3U/2+\Delta$ during the passage procedure in 18 ms. 
In this process, only the configuration $\ket{\ldots11\ldots}$ in the odd-even sites encountered resonance with configuration $\ket{\ldots20\ldots}$. 
Hence, only these two configurations can be adiabatically transferred into each other with high fidelity.
The numerically simulated time-resolved density profile and the corresponding probabilities of atom distributions $\ket{\downarrow,\downarrow,\downarrow,\downarrow,\downarrow,\downarrow}$ and $\ket{\downarrow\downarrow,0,\downarrow\downarrow,0,\downarrow\downarrow,0}$ are plotted in Fig.~\ref{figureS9:adiabatic_passage}(d) and Fig.~\ref{figureS9:adiabatic_passage}(e), respectively.
Numerical simulations indicate the final probability of the atom distribution $\ket{\downarrow\downarrow,0,\downarrow\downarrow,0,\downarrow\downarrow,0}$ exceeds 99.7\%.

Similarly, we employed the same passage procedures as mentioned above for the initial atom distribution $\ket{\downarrow,\uparrow,\downarrow,\uparrow,\downarrow,\uparrow}$, as illustrated in Fig.~\ref{figureS9:adiabatic_passage}(c).
The numerical result indicates that the atom distribution is preserved with a probability exceeding 99.9\%, as presented in Fig.~\ref{figureS9:adiabatic_passage}(f,g).

\textbf{More analysis}.
Despite the high fidelity of the numerical simulation, the actual ramp curves for parameters $J$, $U$, $\delta$, and $\Delta$ exhibited slight deviations from those shown in Fig.~\ref{figureS9:adiabatic_passage}(b) in the experiment.
As a result, the actual fidelity of the adiabatic passage procedure might be lower than shown in Fig.~\ref{figureS9:adiabatic_passage}(e,g).
At this moment, in the first case mentioned above, an atom distribution might exist as $\ket{\ldots,\downarrow,\downarrow,\ldots}$,  making it impossible to realize a transfer from $\ket{\ldots11\ldots}$ to $\ket{\ldots20\ldots}$.
However, these imperfections in the experiment are insignificant, as we added an extra process after the adiabatic passage process.
We then used a spin-dependent superlattice to flip the spin state from  $\ket{\downarrow}$ to $\ket{\uparrow}$ on the right side of the double wells (the even sites shown in Fig.~\ref{figureS9:adiabatic_passage}a), followed by a removal process with a resonant laser beam to clean these atoms in the $\ket{\uparrow}$ state.
At this moment, the total atom number in this chain is no longer equal to the length of the chain; therefore, this sample will be excluded from our final statistics according to our post-selection rules mentioned above.
The same outcomes for infidelity occur in the second case mentioned above.

\section{Calibrations of Bose--Hubbard parameters}

The calibration procedure for the Bose--Hubbard parameters $J$, $U$, $\delta$, and $\Delta$ is identical to those described in our previous works \cite{Zhang2022quantum,Wang2022interrelated}.

\begin{figure*}[ht!]
    \centering     %
    \includegraphics[width=0.99\textwidth]{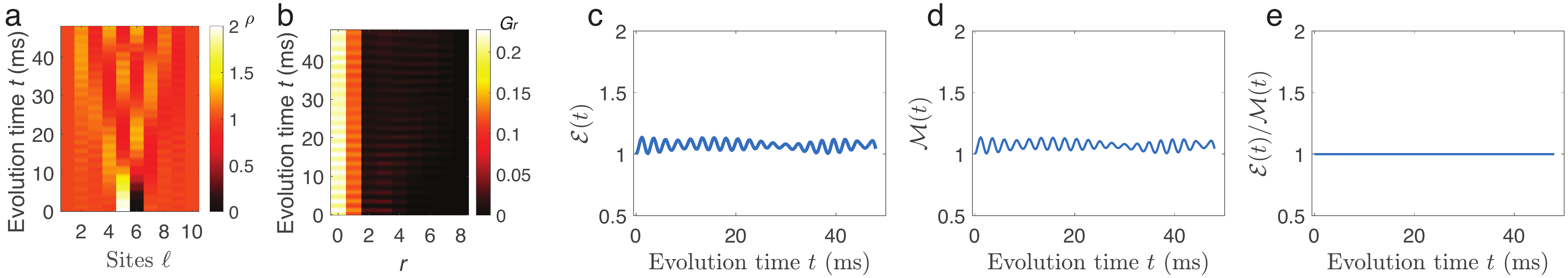}
    \caption{\textbf{Numerical results in the negative mass region}.
    (a) The time-resolved density profiles.
    (b) The time-resolved two-point correlations.
    The extracted time-resolved results of the observables $\mathcal{E}(t)$ (c), $\mathcal{M}(t)$ (d) and $2\mathcal{E}(t)/\mathcal{M}(t)$ (e).
    }
    \label{figureS5:numerical2}
\end{figure*}

\begin{figure*}[ht!]
    \centering     %
    \includegraphics[width=0.92\textwidth]{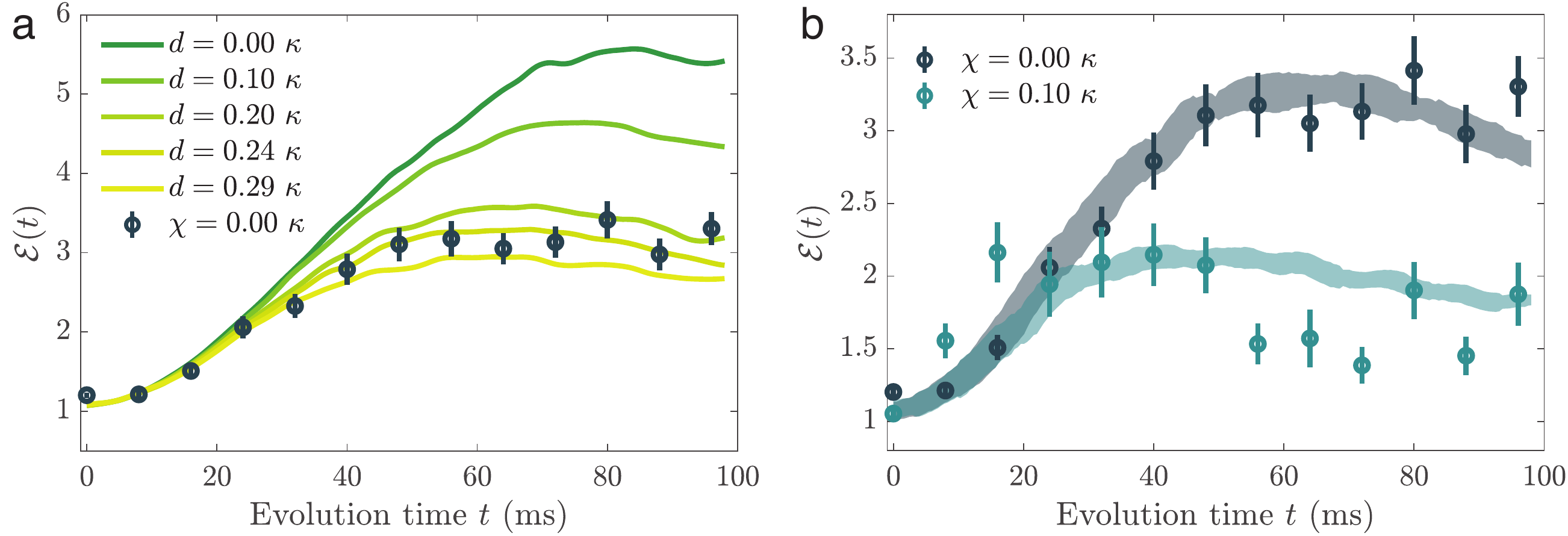}
    \caption{\textbf{Influence of disorder}.
    (a) Numerical simulation results in the presence of disorders of different amplitudes $d$. 
    (b) The dark blue solid dots are the experimental results under $\chi=0$. The light blue solid dots are the experimental results under $\chi/\kappa=0.1$. The dark-shaded region represents the numerical sampling result under $\chi=0$ with uniformly distributed random disorders in the interval (0, $d=0.24\kappa$). The light-shaded region represents the numerical sampling result under $\chi/\kappa=0.1$ with uniformly distributed random disorders in the interval (0, $d=0.24\kappa$). 
    }
    \label{figureS6:disorders}
\end{figure*}

\section{Numerical simulations}

We perform numerical simulations based on the Krylov-subspace method \cite{Sidje1998} in the main text and this supplementary material.

\subsection{Numerical results}

Figure~\ref{figureS4:numerical1} displays the numerical results for the time-resolved density profiles and two-point correlations and the corresponding extracted observables $\mathcal{E}(t)$, $\mathcal{M}(t)$, and $2\mathcal{E}(t)/\mathcal{M}(t)$ in the positive mass region for different $\chi/\kappa$. Meanwhile, Fig.~\ref{figureS5:numerical2} summarizes the numerical results in the negative mass region.

\subsection{Influences of inhomogeneity}

As pointed out in the main text, experimental results exhibit slight deviations compared to numerical simulations. These discrepancies arise from the inherent disorders originating from the overall trapping potential of the experimental platform. We conduct numerical simulations based on the 1D tilted Bose--Hubbard model (Eq.~\eqref{eq:bhm_sup}) using parameters closely matching those in our experiment. We then introduce an additional site-dependent disorder term to the original energy offset term $\epsilon$ in Eq.~\eqref{eq:bhm_sup}, such that $\epsilon \to \epsilon_l = \epsilon + d_l$. Here, $d_l$ represents a uniformly distributed random number within the interval (0, $d$).

We perform the numerical simulation 500 times, incorporating randomly chosen disorders for each given amplitude $d$, and execute numerical sampling at each evolution time. Subsequently, we analyze these numerical samples using the same method employed in our experiment. The summarized results are displayed in Fig.~\ref{figureS6:disorders}. The experimental result aligns with the numerical sampling when $d/\kappa \approx 0.24$.

The primary source of inhomogeneities in our experiment is the repulsive potential projected by the DMD, which compensates for the Gaussian envelopes originating from the lattice beams \cite{Zhang2022quantum}. Further optimization of the projection pattern on the DMD can help smooth the repulsive potential, thereby mitigating the amplitude of disorders and potentially increasing the available size of our quantum simulator by several times.

\end{document}